\documentclass[aps,prb,groupedaddress,reprint,longbibliography]{revtex4-2}

\pdfoutput=1

\usepackage{graphicx}
\usepackage{amsmath,amsfonts,amssymb,mathtools}
\usepackage{color}
\usepackage[dvipsnames]{xcolor}
\usepackage{subfigure}
\usepackage{physics}
\usepackage{dsfont}
\usepackage{hyperref}
\usepackage{booktabs}
\usepackage{url}
\usepackage{leftidx}
\usepackage{textcomp}
\usepackage{gensymb}
\usepackage{rotating}
\usepackage{esvect}
\usepackage{booktabs}
\usepackage{marvosym}
\usepackage{lmodern}
\usepackage{algorithm}
\usepackage{algpseudocode}
\usepackage[version=4]{mhchem}

\definecolor{emerald}{rgb}{0.31, 0.78, 0.47}
\definecolor{blue(ncs)}{rgb}{0.0, 0.53, 0.74}

\hypersetup{
     colorlinks=true,
     linkcolor=blue(ncs),
     filecolor=blue,
     citecolor=emerald,      
     urlcolor =blue(ncs),
     breaklinks=true
}
	    
\definecolor{lightgray}{rgb}{0.9,0.9,0.9}	   
\definecolor{green}{rgb}{0,0.5,0}
\definecolor{red}{rgb}{1,0,0}
\definecolor{blue}{rgb}{0,0,0.5}

\long\def\symbolfootnote[#1]#2{\begingroup%
\def\thefootnote{\fnsymbol{footnote}}\footnotetext[#1]{#2}\footnotemark[#1]\endgroup}

\begin{document}

\title{Game of Life on Archimedean Lattices:\\ Glider Guns and Phase Dynamics}

\author{Henrik S.~Guttesen}
 \email{henrisro@proton.me}
  \affiliation{Danish Fundamental Metrology, 2970 H{\o}rsholm, Denmark}
\date{\today}

\begin{abstract}
I explore Conway's Game of Life (GoL) on six composite Archimedean lattices. On the Kagome lattice, on which small gliders and puffers appear particularly frequently across inputs, I use the output of a symmetry-constrained evolutionary search algorithm to construct a novel glider gun. The glider gun comprises four interacting bouncers and stably emits a small glider every 276th generation. Serving as an extension of classical GoL, I also propose cells with a phase degree of freedom and an associated local phase rule, which on the Kagome lattice is demonstrated to host phase-periodic gliders. This enables the possibility of phase-sensitive and interference-based computations.
\end{abstract}

\maketitle

%
%%
%%%
\section{Introduction}
\label{sec:Intro}
%%%
%%
%
Cellular automata (CA) are discrete models of computation and dynamical systems that trace their origins to the work of Ulam and von Neumann in the 1940s on self-reproduction and the construction of complex systems from simple, locally interacting components~\cite{vonNeumann51, vonNeumann1966}. Von Neumann's work established a general framework for investigating how computation, self-replication, and reliable complex behavior can emerge from networks of simple elements governed by local rules. This perspective has since influenced a broad range of fields, including computational complexity and universality~\cite{UllmanHopcroft79}, information theory and error correction~\cite{MacKay03}, evolutionary and unconventional computation~\cite{Wolfram02, Adamatzky10}, and models of biological and physical systems.

A cornerstone of CA is Conway's Game of Life (GoL), originally conceived as a simple mathematical model of self-organization and the emergence of complex behavior from local interactions reminiscent of biological birth, survival, and death~\cite{Gardner71}. In GoL, a binary-state cell on the square lattice is deterministically updated from one generation to the next according to simple rules governed by its Moore neighborhood (corner- and edge-sharing plaquettes). Remarkably, these local rules strike a delicate balance between order and chaos, giving rise to an extraordinary diversity of emergent structures and dynamics. A profound consequence of this balance is that GoL is Turing complete, implying that
\begin{equation}
\mathrm{Computations} \subseteq \mathrm{GoL~dynamics},
\label{eq:TuringComplete}
\end{equation}
with the computation determined solely by the initial configuration of the cells~\cite{JG2022conway}.

An extensive number of alternative GoL rules have been systematically explored and cataloged~\cite{LifeWiki, Catagolue, Apgsearch}. Numerous other generalizations have also been proposed, including three-dimensional variants~\cite{Bays1987, Bays2006}, continuous GoL~\cite{SmoothLife11, Lenia19, KumarEA24}, and quantum GoL~\cite{BlehEA12, ArrighiEA10, EscanezEA26}. Furthermore, possible physical realizations of CA based on interacting particles, as well as programmable optical implementations, have been demonstrated~\cite{ReichthardtEA03, ZhangEA24}.

The majority of the reported GoL generalizations and alternative rules retain the square-lattice geometry, with all cells having coordination number $z_{\mathrm{square}} = 8$ under the Moore neighborhood. Other studies have considered triangular, pentagonal, hexagonal, and quasiperiodic lattices~\cite{Adamatzky10}, as well as, more sporadically composite and other non-square geometries~\cite{CodeGolf}. To the best of my knowledge, among the reported two-state GoL-like rules on non-square lattices, the triangular lattice is the only one, besides Conway's original square-lattice rule, for which a glider gun has been demonstrated~\cite{Bays2007TheDO}. 

A glider gun is a self-sustaining and oscillating structure that periodically generates and releases gliders, producing an unbounded stream of mobile patterns without being consumed in the process. Glider guns are important because they enable controlled signal generation and transport, making them key components for constructing logical circuits, information-processing systems, and computationally universal CA. Although glider guns are absent from many cellular-automaton rules, some dynamically rich rules support them abundantly, even allowing their emergence from random initial conditions~\cite{Martinez2010, MartinezEA2014, Seeds}. Their existence is nevertheless an important dynamical and computational feature, as it demonstrates the ability of a rule to support mechanisms for persistent information generation and collision-based computation~\cite{Adamatzky02}.

\begin{table*}
\centering
\caption{Summary of life forms discovered under the B3/S23 rule on six Archimedean lattices. Searches combined random initial conditions with manual and evolutionary exploration. $^\star$Automatic oscillator detection was unreliable because random configurations on this lattice frequently grow without bound.}
\label{tab:LifeForms}
\begin{tabular}{l c l c c c}  
\toprule
Lattice (face configuration) & Still life & Oscillator (period) & Glider & Puffer & Glider gun \\ \midrule
Truncated square $(4,8^2)$ & \checkmark & \checkmark \; (2, 4, 5, 6, 7, 14) & - & - & - \\
Kagome $(6,3,6,3)$ & \checkmark (topo) & \checkmark \; (2,3,4,5,7,8,9,10,32) & \checkmark & \checkmark & \checkmark \\
Maple leaf $(3^4,6)$ & \checkmark & \checkmark \; (2, 3, 8, 20, 24) & - & - & - \\
Star $(3,12^2)$ & \checkmark (topo) & \checkmark \; (2,3,4,5,8,9,18) & - & - & - \\
Truncated trihexagonal $(4,6,12)$ & \checkmark (topo) & \checkmark \; (2,3,5,6,9,15) & - & - & - \\
Snub square $(3^2,4,3,4)$ & \checkmark & \checkmark \; $(2)^\star$ & \checkmark & - & - \\ 
\bottomrule
\end{tabular}
\end{table*}

Here, I embed Conway's GoL rule on six composite Archimedean lattices and report findings in terms of structures such as oscillators, gliders, and puffers, primarily using randomized searches. On the Kagome lattice, of which small gliders and puffers appear particularly frequently across inputs, an evolutionary algorithm is applied to discover a bouncer, which in turn is used to construct a novel glider gun comprising four interacting bouncers, stably emitting a small glider every 276th generation. Serving as an extension of classical GoL, I also propose extending cells with a local phase degree of freedom and an associated local phase rule, which on the Kagome lattice may be tuned to produce coherent configurations such as phase-periodic gliders.

\section{Game of life on Archimedean lattices formalities}
\label{sec:Kagome}
On a given lattice, $L$, I define a configuration as a collection of binary cell values assigned to the plaquettes of the lattice at generation $n$: $\mathcal{C}^{(n)}_L = \lbrace c_{p}^{(n)} : c_{p}^{(n)} \in \lbrace 0, 1 \rbrace \rbrace_{p}$ ($p$ indexing the plaquettes of the lattice). Plaquettes of value 0 (respectively 1) are referred to as dead (respectively living) cells. Conway's GoL rule may be defined by the (deterministic) iteration
\begin{equation}
    c_{p}^{(n+1)}= 
    \begin{cases}
        1 &\mathrm{if}\; c_{p}^{(n)} = 0 \; \land \; \sum_{p' \in M(p)} c_{p'}^{(n)} \in B \\
        1 &\mathrm{if} \; c_{p}^{(n)} = 1 \; \land \; \sum_{p' \in M(p)} c_{p'}^{(n)} \in S \\
        0 &\mathrm{otherwise}
    \end{cases},
    \label{eq:GoLrule}
\end{equation}
where $M(p)$ denotes the set of plaquettes in the Moore neighborhood (corner- or edges-sharing plaquettes) of $p$, and $B$ and $S$ contain numbers defining the birth and survival of rules for cells. In conventional square lattice GoL $B = \lbrace 3 \rbrace$ and $S = \lbrace 2, 3\rbrace$, usually referred to as the ``B3/S23'' rule. In the literature and online GoL communities, however, a vast variety of other rules have been explored numerically, see e.g.~Ref.~\cite{Catagolue}. 

\begin{figure}[b!th]
	\centering
    \includegraphics[width=0.5\linewidth]{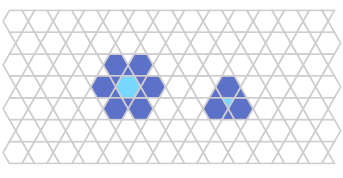}
    \caption{Moore neighborhood (violet) of a hexagonal and a triangular plaquette (turquoise) on the Kagome lattice, comprising twelve and six edge- and corner-sharing neighboring plaquettes, respectively.}
	\label{fig:Neighborhood}
\end{figure}

The Archimedean lattices may be defined as the eleven 2D planar tilings composed of regular polygons such that all vertices are equivalent (surrounded by the same sequence of polygons). In other words, on the Archimedean lattices, each vertex has the same coordination number. The same property does not hold true for the vertices on Laves tilings dual to the eight composite Archimedean lattices (tilings composed of more than one regular polygon type). The Kagome lattice, for example, comprises hexagonal and triangular plaquettes with coordination numbers (in a Moore-neighborhood-sense) $z_{\mathrm{hex}} = 12$ and $z_{\mathrm{tri}} = 6$, respectively, see Fig.~\ref{fig:Neighborhood} (the average Kagome coordination number, $\bar{z}_{\mathrm{Kagome}} = \frac{1}{4} z_{\mathrm{hex}} + \frac{3}{4} z_{\mathrm{tri}} = 15/2$, is, however, notably close to the square lattice value, $z_{\mathrm{square}} = 8$). Hence, on the composite Archimedean lattices, straightforwardly embedding the ``B3/S23'' rule implicitly implies plaquette-dependent rules relative to the coordination number.

\subsection{Summary of life forms and search strategy}

This section provides a summary of the discovered life forms after embedding Conway's GoL rules on six Archimedean lattices (focusing on the ones not already investigated in Ref.~\cite{CodeGolf}), employing a strategy combining random and evolutionary searches with manual modifications and bookkeeping of the most interesting configurations. Table~\ref{tab:LifeForms} summarizes the life forms discovered with the B3/S23 rule on the six (composite) Archimedean lattices. 

The employed search strategy combines a stochastic automated exploration with interactive human-guided refinement for discovering structured patterns in GoL on arbitrary lattices (building on \href{https://codegolf.stackexchange.com/users/194/peter-taylor}{Peter Taylor}'s abstract lattice construction in Ref.~\cite{CodeGolf}). The primary search is performed by generating a random (or optionally user-provided) starting configuration, simulating its dynamical evolution for a number of generations, and evaluating them against objective criteria (e.g., structural recovery, desired output, and minimal debris) condensed in a \emph{fitness} score. The search proceeds by generating local (or optionally symmetrized) mutations of the starting configuration having achieved the highest fitness score so far and repeating the previous steps. This process may be used to identify structures with particular properties such as reflectors or bouncers, depending on the designed fitness function.

After searches, configurations of interest may be transferred to a graphical user interface (GUI) to perform fine-grained, local edits and perturbations through direct manipulation, with immediate visual  feedback from the underlying automaton dynamics. The GUI supports stepwise evolution, state inspection, and spatially bounded exploration (such as via an absorbing radius), enabling efficient validation, stabilization, and optimization of candidate patterns. This hybrid approach proved efficient to locate extremely rare configurations by brute-force evolution tweaked by human intuition across lattices.

\subsection{Still life: topological strings}
\label{sec:StillLife}
\begin{figure}[h!tb]
	\centering
	\includegraphics[width=0.6\linewidth]{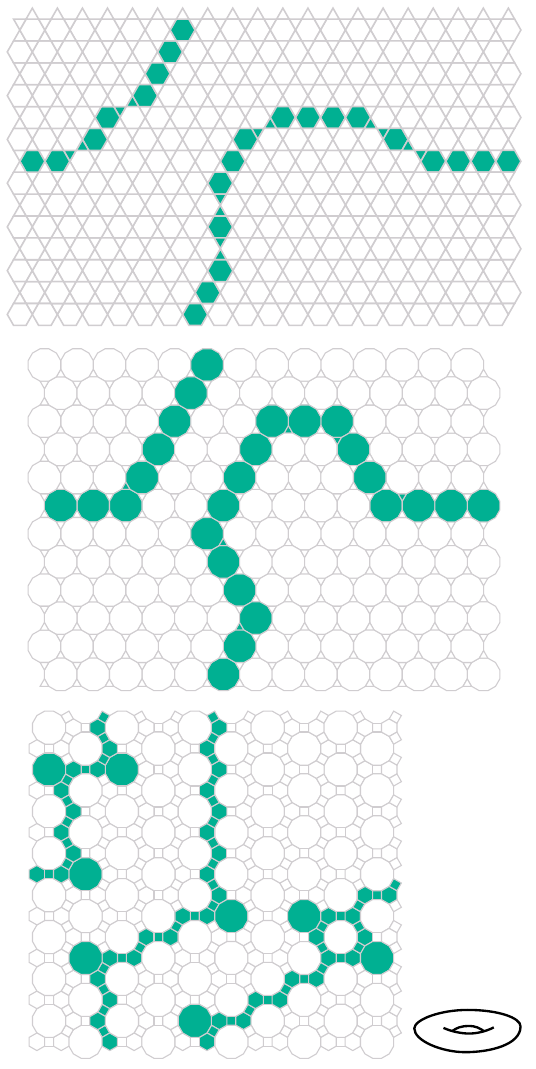}
    \caption{Topological still life on Archimedean lattices $(3,6,3,6)$, $(3,12^2)$, and $(4,6,12)$ from top to bottom with periodic boundary conditions.}
	\label{fig:topostrings}
\end{figure}

Considering first the case of static life forms (``still life'') under the rule of Eq.~\eqref{eq:GoLrule}, it is possible to construct arbitrarily long, meandering strings of still life on Archimedean lattices $(3,6,3,6)$, $(4,6,12)$, and $(3,12^2)$. On these lattices in particular, the strings of still life may form along (and connect between) three independent directions, allowing the strings to wind topologically non-trivially if the lattice is embedded on a higher-genus surface; see Fig.~\ref{fig:topostrings}.

While such topological strings in themselves are vulnerable to (and usually collapse by) the impact of gliders or dynamic life forms, they open up the possibility of a new class of robust memory states: a bridge between CA and topological defect theory and a route to computation where information is stored and manipulated by the connectivity and winding of extended living strings rather than by isolated gliders or still lives.

\subsection{Oscillators}
\label{sec:Oscillators}
Oscillators are dynamic life forms that repeat after a certain number of generations, referred to as the period of the oscillator. Oscillators are fundamental to the computational and structural richness of GoL, as they can act as clockwork, memory, and timing primitives. Depending on context and form, they can both emit, absorb, reflect, or gate moving patterns. 

\begin{figure}[b!th]
	\centering
	\includegraphics[width=\linewidth]{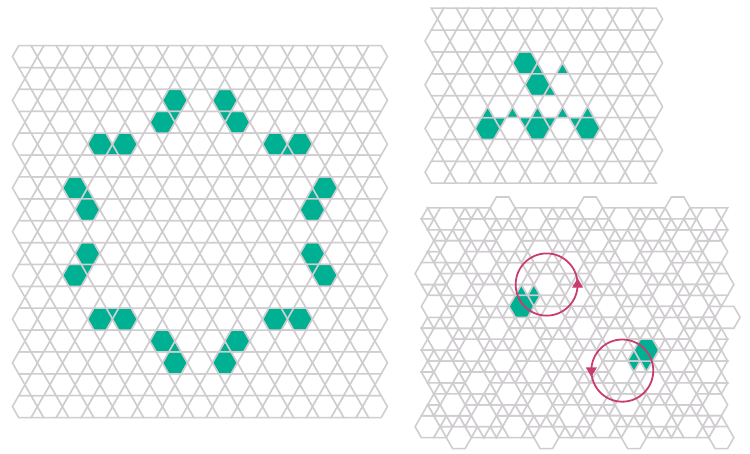}
    \caption{Oscillators on depleted traingular lattices. Left: star-shaped oscillator of period 3 on the Kagome lattice. Top right: Period-32 oscillator on the Kagome lattice, reminiscent of a bouncing glider. Lower right: period-24 oscillators comprising chiral circulating configurations on the maple leaf lattice.}
	\label{fig:oscillators}
\end{figure}

In the investigated composite Archimedean lattices, oscillators are ubiquitous; short-period oscillators, in particular ``blinkers'' (period-2 oscillators), are found within seconds of random searches on all lattices. Examples of both longer-period oscillators and more peculiar oscillator shapes on the Kagome and Maple leaf lattice are shown in Fig.~\ref{fig:oscillators}.

\subsection{Gliders and puffers}
\label{sec:Gliders}
A glider in GoL is a finite pattern that repeats its shape after a fixed number of generations while undergoing a nonzero spatial translation, i.e., it is a moving periodic solution of the automaton. A puffer, on the other hand, is a finite pattern that translates across the lattice while emitting debris (either persistent or chaotic), meaning that in a comoving frame the configuration is not eventually periodic.

\begin{figure}[t!bh]
	\centering
	\includegraphics[width=\linewidth]{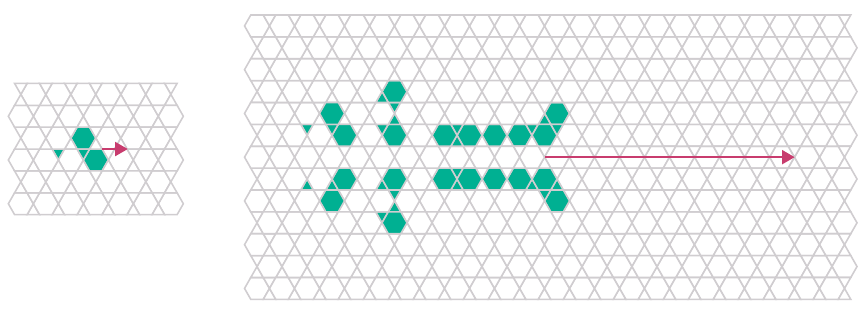}
    \caption{Kagome lattice GoL gliders with red translation vectors indicating the translation before a repetition of the pattern occurs (known as the glider period). Left: small glider, $g$, of period 4. Right: big $\mathrm{C}_2$ symmetric glider, $G$, of period 40 comprising an accompanying trail of two small gliders that wipe out intermediate debris.}
	\label{fig:Kagomegliders}
\end{figure}

\begin{figure}[b!th]
	\centering
	\includegraphics[width=0.45\linewidth]{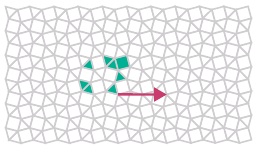}
    \caption{Snub square lattice GoL glider of period 8.}
	\label{fig:Snubsquareglider}
\end{figure}

\begin{figure}[t!bh]
	\centering
	\includegraphics[width=0.70\linewidth]{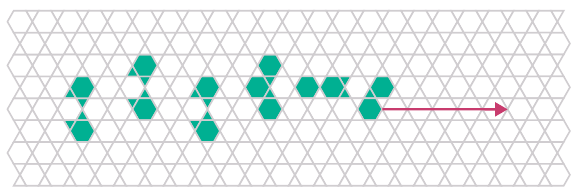}
    \caption{Puffer on the Kagome lattice, leaving behind a predictable trail of still lives (footsteps) as it traverses the lattice. The mere existence of this puffer demonstrates infinite (yet sparse) growth on this lattice.}
	\label{fig:puffer}
\end{figure}

Among the six investigated lattices, gliders were identified on two: the Kagome lattice and the Snub square lattice, see Fig.~\ref{fig:Kagomegliders} and \ref{fig:Snubsquareglider}. On the Kagome lattice, a puffer (with clean debris) is obtained by a simple modification of the small glider, see Fig.~\ref{fig:puffer}.

\subsection{Bouncer walls and bouncers}
\label{sec:Bouncers}
A bouncer wall is a finite static pattern that, upon interaction with an incoming propagating excitation (e.g., a glider), returns to its initial configuration while producing an outgoing excitation traveling in a different direction, i.e., it is a periodic scattering structure. A bouncer may then be defined as a finite pattern composed of one or more bouncer walls that repeatedly redirects a propagating excitation along a closed trajectory, such that the excitation periodically revisits the walls and reconstructs the original configuration. Equivalently, a bouncer is nothing but a particular type of oscillator comprising a recurring moving excitation sustained by repeated interactions with bouncer walls.

Automated and evolutionary search methods have previously been applied to the discovery of localized structures in CA, including gliders and glider guns~\cite{Wuensche2002}. In particular, genetic and stochastic algorithms have been used to evolve initial configurations or rule parameters toward persistent propagating structures and periodically emitting configurations~\cite{MitchellEA93, SapinEA04, SapinEA07, SapinEA08, SapinEA10}. This demonstrates the utility of evolutionary search for navigating the large configuration spaces of CA.

\begin{figure}[tb]
	\centering
	\includegraphics[width=0.8\linewidth]{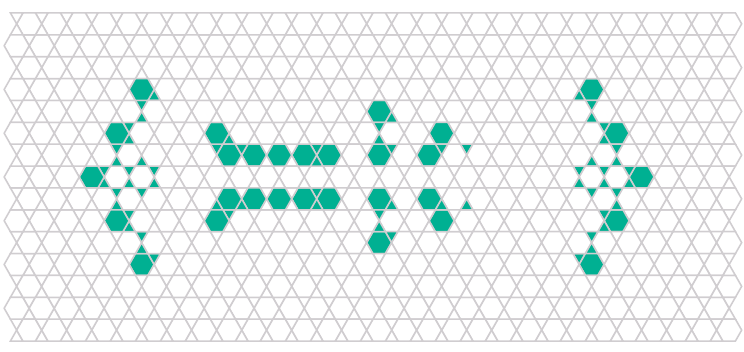}
    \caption{A ($\mathrm{C}_2$ symmetric) bouncer on the Kagome lattice comprising two bouncer walls and a big bouncing glider, $G$. The distance between the bouncer walls can be trivially expanded to encompass further glider periods to yield any bouncer period of the form $B_n = 76 + 40n$, for any $n \in \mathbb{N}$. Here $n = 1$.}
	\label{fig:Kagomebouncer}
\end{figure}

\begin{algorithm}[H]
\caption{Symmetry-constrained evolutionary search for a GoL bouncer wall.}
\label{alg:Bouncer}
\begin{algorithmic}[1]
\Require Static seed structure $S$, incoming glider $G_{\mathrm{in}}$, desired output pattern $G_{\mathrm{out}}$, symmetry axis $l$
\State Initialize best fitness $F_{\mathrm{best}} \gets -\infty$
\State Initialize best candidate $S_{\mathrm{best}} \gets S$

\While{termination criterion not met}
    \State $S' \gets S$
    \Comment{Apply local symmetric mutations}
    \For{$k = 1,\dots,N_{\mathrm{mut}}$}
        \State Select anchor cell $a \in S'$
        \State Select nearby cell $c$ within radius $R$ of $a$
        \State Compute reflected cell $c^\star = \mathrm{Reflect}(c,l)$
        \State Toggle occupancy of $c$ in $S'$
        \If{$c^\star \neq c$}
            \State Toggle occupancy of $c^\star$ in $S'$
        \EndIf
    \EndFor
    \Comment{Simulate collision dynamics}
    \State $X_0 \gets S' \cup G_{\mathrm{in}}$
    \State Evolve $X_0$ under GoL dynamics until extinction, recurrence, or $T_{\max}$ generations
    \Comment{Evaluate candidate}
    \State Compute structural recovery score $f_{\mathrm{static}}$
    \State Compute target overlap score $f_{\mathrm{out}}$
    \State Compute debris penalty $p_{\mathrm{extra}}$
    \State Compute fitness
    \[
    F = \left(\tfrac12 f_{\mathrm{static}} + \tfrac12 f_{\mathrm{out}}\right)(1-p_{\mathrm{extra}})
    \]
    \If{$F > F_{\mathrm{best}}$}
        \State Archive candidate configuration
        \State $F_{\mathrm{best}} \gets F$
        \State $S_{\mathrm{best}} \gets S'$
    \EndIf
\EndWhile

\State \Return optimized bouncer wall $S_{\mathrm{best}}$
\end{algorithmic}
\end{algorithm}

\begin{figure*}[t!b]
	\centering
	\includegraphics[width=\linewidth]{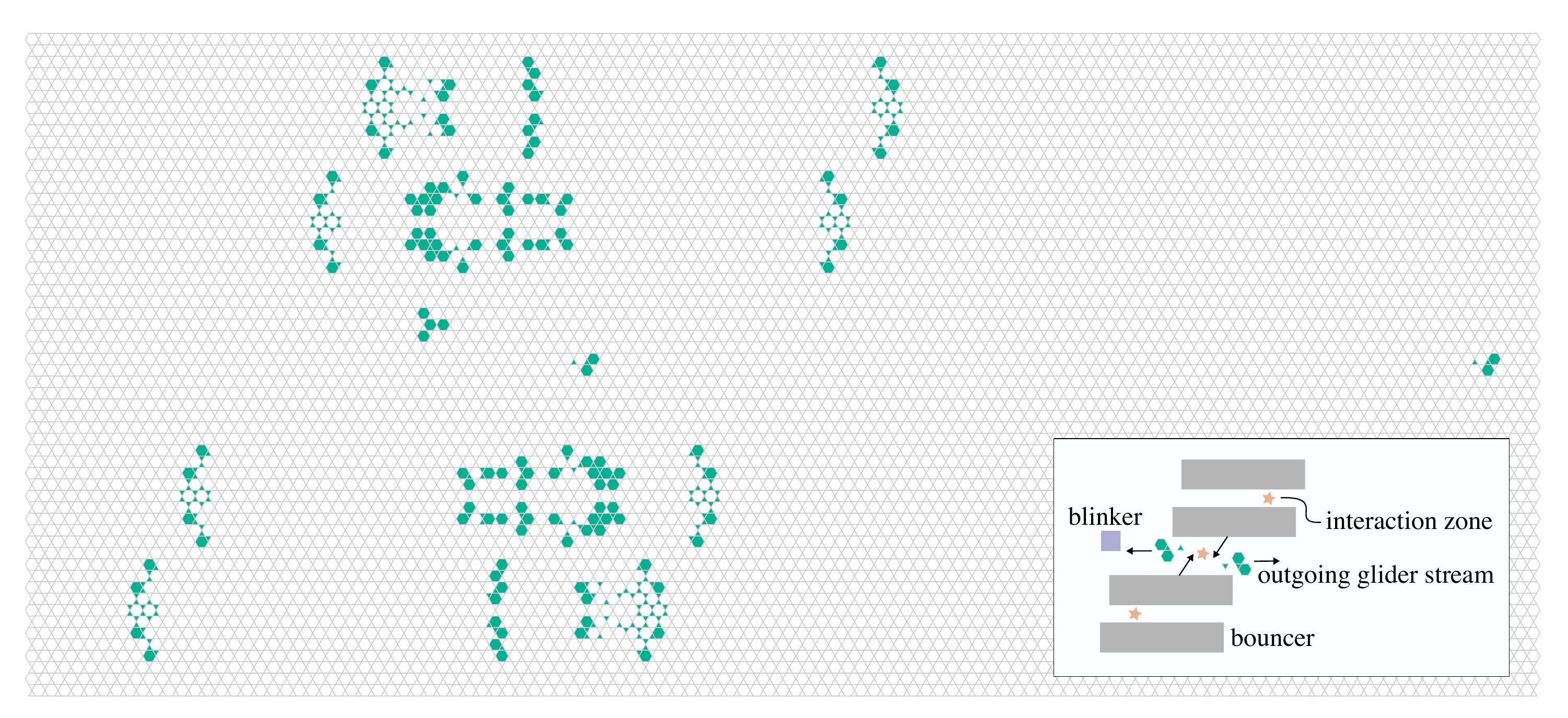}
    \caption{Glider gun construction of period 276 with a single emitting stream of gliders (schematic illustration in the lower right corner). A pair of $B_2$ bouncers are displaced to interact at a single point during every full bouncer period, causing the stable emission of a puffer. A mirror symmetric partner construction ensures that two emitted puffers collide and emit two small gliders, one of which is (optionally) absorbed by a blinker, and one sliding off to the right.}
	\label{fig:glidergun}
\end{figure*}

Here, I adopt a related stochastic-search strategy for a fixed rule on non-square lattices. The searches evolves the initial configuration itself rather than the local update rule, with a symmetry constraint and fitness function tailored to the discovery of a bouncer. The algorithm is summarized in Algorithm~\ref{alg:Bouncer}. Starting from a seed static bouncer wall candidate and a fixed incoming glider, the algorithm repeatedly generates candidate bouncer walls by applying a small number of local random mutations (cell flips within a bounded neighborhood), while enforcing exact mirror symmetry about a prescribed lattice axis. Each candidate is then evolved under the cellular automaton dynamics until extinction, recurrence, or a fixed generation cutoff.

Fitness is evaluated by balancing three objectives: (i) maximal recovery of the mutated static structure after collision (structural persistence), (ii) maximal overlap with a prescribed target output pattern (desired scattering channel), and (iii) penalization of excess live cells (spurious debris). Candidates improving the best-known fitness are archived, visualized, and adopted as new candidate solutions; iteration continues until a desired fitness score is reached. In effect, the method performs a localized, symmetry-preserving evolutionary search in configuration space.

On the Kagome lattice, extensive searches spanning on the order of $\mathcal{O}(10^5)$ sets of mutations, starting from a handpicked initial configuration, resulted in a bouncer wall (tailored for the big glider $G$, owing to its long period), supporting a family of bouncers with period of the form $B_n = 76 + 40n$ for any $n \in \mathbb{N}$, see Fig.~\ref{fig:Kagomebouncer}.

\subsection{Glider gun}
\label{sec:GliderGun}
A glider gun is a finite pattern that periodically emits gliders while returning to its original configuration after a fixed number of generations, i.e., it is a localized periodic solution that acts as a persistent source of gliders. 

The bouncer family of Sec.~\ref{sec:Bouncers} provides the key ingredient to construct a glider on the Kagome lattice. A logical step towards this goal concerns making the intermediate debris of two big gliders, each belonging to a separate bouncer, interact to produce an output glider. While one can get tantalizingly close to this construction using two appropriately displaced $B_1$ bouncers (see Appendix~\ref{app:Nearmiss} for a snapshot of the near-miss configuration), exhausting the displacement options for two $B_1$ bouncers yields no success. Moving on to using $B_2$'s, however, a particular displacement option was found to produce an outgoing puffer. Combining this observation with a particular puffer collision that results in two outgoing gliders (erasing all puffer debris in the process), the basic ingredients are in place to construct a glider gun. 

The final result is shown in Fig.~\ref{fig:glidergun}; the construction comprises four bouncers (and in this case optionally a blinker to absorb $g$ gliders in one outgoing direction). Two big bouncing gliders, $G$, are carefully phase shifted and displaced to interact at a single point during every full bouncer period, causing the stable emission of a puffer as the intermediate debris of the interaction zone clears. A mirror symmetric partner construction ensures that two emitted puffers interact to emit two small gliders $g$, one of which is (optionally) absorbed by a blinker~\footnote{A two-sided glider gun, sending off gliders both left and right, is obtained by removing the blinker.}, and one sliding off to infinity.

The discovery of this glider gun demonstrates that the dynamical richness of the Kagome Game of Life extends beyond isolated mobile patterns to include self-sustaining structures capable of periodically generating propagating information carriers. Although glider guns are comparatively rare in CA, their existence is of particular interest because they provide a persistent source of gliders, which constitute the fundamental building blocks for many collision-based computational constructions. More broadly, the mechanism underlying the present construction highlights how synchronized interactions between composite mobile structures can give rise to qualitatively new dynamical behavior, suggesting that the space of complex objects and interaction mechanisms on Archimedean lattices remains only partially explored. \\

%
%%
%%%
\section{Multi-component Game of Life: Phase interactions}
\label{sec:Phaseinteractions}
%%%
%%
%
Serving as a generalization of classical GoL, consider cells parametrized by a binary state value $r_p^{(n)} \in \lbrace 0,1 \rbrace$ \emph{and} a phase $\phi_p^{(n)} \in [-\pi, \pi)$: $c_p^{(n)} = r_p^{(n)}\mathrm{exp} (i\phi_p^{(n)}) $. We seek a generalization of classical GoL in which the phase serves as an internal state that evolves depending on the classical GoL dynamics, so that the generalized GoL becomes a multi-component CA. For simplicity, I consider the following generalization of the rule of Eq.~\eqref{eq:GoLrule} in which cells are born with zero phase, and in which a cell's phase depends upon the Moore neighborhood phase sum and a self-rectification controlled by the phase interaction parameter $\kappa$:

\begin{widetext}
\begin{equation}
    c_{p}^{(n+1)}= 
    \begin{cases}
        1 &\mathrm{if}\; r_{p}^{(n)} = 0 \; \land \; \sum_{p' \in M(p)} r_{p'}^{(n)} \in B \\
        \exp\left(i\left[\sum_{p'\in M(p)} \phi_{p'}^{(n)}- \kappa \phi_{p}^{(n)}\right] \right) &\mathrm{if} \; r_{p}^{(n)}  = 1 \; \land \; \sum_{p' \in M(p)} r_{p'}^{(n)} \in S \\
        0 &\mathrm{otherwise}
    \end{cases}.
    \label{eq:PhaseRule}
\end{equation}
\end{widetext}
By construction, whether a cell is born, dies, or survives from one generation to the next is not affected by the cell's phase. Instead, a cell's phase is updated solely based on the local phase environment (and as such affected by neighboring cells more than one generations old). 

\subsection{Static periodic strings of life}
\label{sec:Toeplitz}
\begin{figure}[h!tb]
	\centering
    \includegraphics[width=0.9\linewidth]{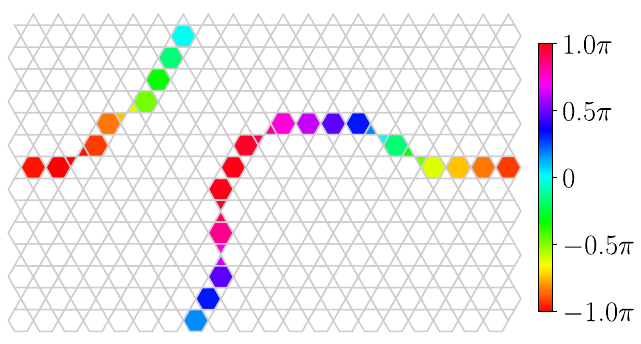}
\caption{Example of phase-decorated string life (configuration proportional to $\boldsymbol{\theta}_1$) with periodic boundary conditions.}
	\label{fig:Phasestring}
\end{figure}

The phase dynamics of Eq.~\eqref{eq:PhaseRule} results in tractable solutions for still lives in which each cell has precisely two living neighbors owing their circulant Toeplitz phase iteration. Such solutions may for example be topological string lives on the Kagome lattice (see, e.g., the top panel of Fig.~\ref{fig:topostrings}). For these string lives the phase evolution on the living cells is governed by the iteration
\begin{equation}
    \boldsymbol{\phi}^{(n+1)} = T\boldsymbol{\phi}^{(n)},
\end{equation}
where $\boldsymbol{\phi}^{(n)}\in\mathbb{T}^N$ denotes the vector of phases at generation $n$, interpreted modulo $2\pi$, and $T$ is the nearest-neighbor Toeplitz matrix
\begin{equation}
    T = \begin{bmatrix}
    -\kappa & 1 & 0 & 0 & \dotsb & 0 & 1
    \\
    1 & -\kappa & 1 & 0 & \dotsb & 0 & 0 \\
    0 & 1 & -\kappa & 1 & \dotsb & 0 & 0 \\
    \vdots & & & & \ddots & & \vdots \\
    0 & 0 & 0 & 0 & \dotsb & -\kappa & 1 \\
    1 & 0 & 0 & 0 & \dotsb & 1 & -\kappa 
    \end{bmatrix},
\end{equation}
in a basis consecutively ordered along the string of living cells. The solution is formally
\begin{equation}
\boldsymbol{\phi}^{(n)} = T^n\boldsymbol{\phi}^{(0)},
\label{eq:Iteration}
\end{equation}
with all components are defined modulo $2\pi$. With periodic boundary conditions, $T$ is circulant and is diagonalized by the discrete Fourier basis. The eigenvalues are~\cite{Noschese2013} 
\begin{equation}
\lambda_k = 2\cos\left(\frac{2\pi k}{N}\right)-\kappa,
\qquad
k=0,\ldots,N-1,
\label{eq:Eigenvalues}
\end{equation}
with corresponding eigenvector components $\theta_{k,j} =\sin(\frac{2 \pi k j}{N})$, see Fig.~\ref{fig:Phasestring}. Here, $N$ is the number of living cells in the string, indexed by $j$. Expanding the initial phase configuration as $\boldsymbol{\phi}^{(0)}=\sum_k c_k\boldsymbol{\theta}_k$ yields
\begin{equation}
\boldsymbol{\phi}^{(n)}=\sum_k c_k\lambda_k^n\boldsymbol{\theta}_1
\pmod{2\pi},
\end{equation}
showing that each spatial Fourier mode evolves independently.
\begin{figure*}[h!tb]
	\centering
    \includegraphics[width=0.9\linewidth]{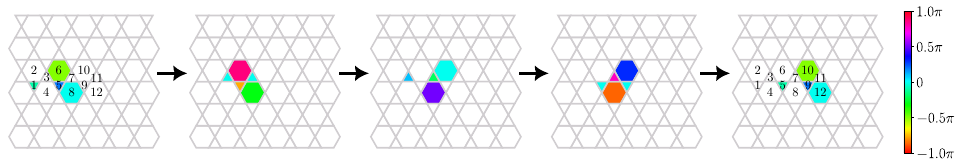}
\caption{Phase-periodic glider corresponding to Eq.~\eqref{eq:glider0} with $\alpha = \pi/5$, $\beta = 5\alpha/2$, $\gamma = -\alpha/2$, and $\kappa = 4/3$.}
	\label{fig:Phaseglider}
\end{figure*}

The modulo-$2\pi$ nature of the phase variables fundamentally alters the asymptotic dynamics compared to a linear evolution on $\mathbb{R}^N$. Rather than diverging or decaying indefinitely, each eigenmode generates an orbit on the torus $\mathbb{T}^N$, whose character is determined by the corresponding eigenvalue. Exact stationary and period-two modes occur for $\lambda_k=1$ and $\lambda_k=-1$, respectively, corresponding to the discrete parameter values
\begin{equation}
\kappa = 2\cos\left(\frac{2\pi m}{N}\right)\mp1,
\qquad
m=0,\ldots,\left\lfloor\frac{N}{2}\right\rfloor.
\end{equation}
Equivalently, these modes exist precisely when
\begin{equation}
\frac{N}{2\pi}\arccos\left(\frac{\kappa\pm1}{2}\right)\in\mathbb{Z}.
\end{equation}
For all remaining values of $\kappa$, the phase evolution is generally aperiodic, with quasiperiodic or ergodic-like trajectories expected for generic initial conditions. The interaction parameter $\kappa$ therefore acts as a spectral tuning parameter that selectively locks particular Fourier wavelengths into coherent periodic evolution while driving others toward increasingly complex dynamics on the phase torus. This suggests that static strings may simultaneously support multiple classes of phase excitations, providing a natural mechanism for encoding and processing information beyond the binary cellular state.

\subsection{Phase periodic gliders}
\label{sec:Phaseperiodic}

I now consider dynamic life forms under the multi-component CA and demonstrate the existence of simultaneously space- and phase-periodic solutions. 

Consider a glider ansatz with three phase degrees of freedom $(\alpha, \beta, \gamma)$,
\begin{equation}
g^{(0)} = \lbrace e^{i\gamma}_1, e^{i\alpha}_5, e^{-i\beta}_6, 1_8, \boldsymbol{0} \rbrace,
\label{eq:glider0}
\end{equation}
where the labels index the plaquettes positions shown in Fig.~\ref{fig:Phaseglider}. Evolving this glider by one period using the phase rule of Eq.~\eqref{eq:PhaseRule} results in
\begin{equation}
\begin{aligned}
g^{(4)} &= \lbrace e^{3(\alpha-\beta)(\kappa - 1)^2}_5, e^{3i(\alpha-\beta)(2\kappa-1)(\kappa-1)}_{9}, \\
&\hspace{10pt}e^{3i(\alpha-\beta)(\kappa-1)(\kappa-2)}_{10}, 1_{12}, \boldsymbol{0} \rbrace.
\end{aligned}
\label{eq:glider4}
\end{equation}
Requiring $g^{(4)}$ to be phase periodic with $g^{(0)}$ yields two solutions, each parametrized by \emph{one} phase degree of freedom:
\begin{equation}
    (\alpha, \beta, \gamma; \kappa) = \left( \alpha, \frac{\alpha}{4},\frac{\alpha}{4}; \frac{2}{3} \right) \lor \left( \alpha, \frac{5\alpha}{2},-\frac{\alpha}{2}; \frac{4}{3} \right).
\end{equation}
Thus for the ansatz of Eq.~\eqref{eq:glider0}, if the self-interaction of Eq.~\eqref{eq:PhaseRule} is tuned to one of two particular values, the glider of the generalized CA may be phase-periodic. An example for $\kappa = 4/3$ is illustrated in Fig.~\ref{fig:Phaseglider}.

Phase periodic gliders in multi-component CA demonstrate the existence of mobile patterns acting as coherent carriers of both spatial and internal (phase) degrees of freedom. These patterns may be used to encode information not only in the presence of the trajectory of the glider, but also in its phase (or relative phase with respect to other excitations), enabling multi-valued and phase-sensitive computations. Such solutions are intrinsically synchronized signals (because the phase is periodic in the co-moving frame), allowing for controlled interactions in which collision outcomes depend on phase alignment. This may be useful in phase-gated logic, signal multiplexing, robust information transport, or interference-based computation. 

More generally, phase-periodic gliders may be viewed as discrete analogs of coherent excitations with internal modes and as such representing CA with dynamical internal structure in addition to spatial organization. Such generalizations bear their motivation from and resemblance with quantum cellular automata (QCA), in which cells become qubits~\cite{BlehEA12, ArrighiEA10, EscanezEA26}. The approach taken here serves as a middle ground between classical and fully quantum approaches, circumventing the unitarity (irreversibility) issue of Conway's rule encountered in any genuine quantum approach. In the approach presented here, cells carry continuous phase information but the update rules remain deterministic.

%
%%
%%%
\section{Conclusions and Outlook}
\label{sec:Conclusion}
%%%
%%
%
In this paper, I have demonstrated rich dynamical phenomena in Conway's Game of Life on Archimedean lattices, whose mixed coordination numbers naturally generalize the Moore neighborhood of the square lattice. Focusing primarily on the Kagome lattice, where gliders and puffers arise frequently from random initial configurations, I constructed a novel period-276 glider gun comprising four bouncers. The underlying bouncer component was discovered using a $\mathrm{C}_2$ symmetry-constrained evolutionary search algorithm.

In addition, a phase-extended formulation of GoL was introduced incorporating nearest-neighbor phase interactions between living cells. This multi-component cellular automaton was shown to support both periodic and aperiodic phase dynamics in static periodic strings of still life, including topologically non-trivial strings on toroidal embeddings. Moreover, suitable choices of the phase coupling yield fully phase-periodic gliders whose internal phase configuration is restored after each translational period. Such objects suggest the possibility of enriching CA with an additional dynamical degree of freedom capable of carrying information independently of the binary cell state, potentially enabling phase-sensitive logical operations or interference-based computational schemes. These ideas may also provide inspiration for future physical realizations in programmable optical or photonic arrays, where coherent phase is an intrinsic degree of freedom and local interactions may be engineered~\cite{ZhangEA24}.

One possibly interesting direction going forward concerns the construction of quantum Hamiltonians with ground states encoding life-like CA on Archimedean lattices and as such provide a conceptual bridge between computational decidability of the CA and spectral properties of the Hamiltonian. For example, the Feynman--Kitaev construction may be used to build a local Hamiltonian on a chain of qubits whose (``history'') ground state encodes a Life computation~\cite{Kitaev02}. 

\begin{acknowledgments}
I thank Andrew Adamatzky, Freja Schou Guttesen, and Zhao Zhang for providing useful comments and connections. 
\end{acknowledgments}

%
%%
%%%
\section*{Generative AI disclaimer}
%%%
%%
%
Development of the evolutionary Algorithm~\ref{alg:Bouncer}, as well as GUI and general code development, was carried out using ChatGPT 5.2 in incremental steps followed by human testing and verification. The glider gun construction logic and the input for the evolutionary searches were developed by manual experimentation and without the use of AI. Some text sections herein were improved by formulations and suggestions resulting from specific prompts to ChatGPT 5.2. 

%
%%
%%%
\section*{Data and code availability}
%%%
%%
%
Java code for lattice constructions, random and symmetry-constrained evolutionary searches, an interactive GUI, oscillators of Table~\ref{tab:LifeForms}, and a collection of the most interesting configurations, including particular glider collisions, identified during the searches, are available from my \href{https://github.com/henrisro313/Games_of_life/tree/main}{GitHub repository}.

\appendix

\section{Near-miss glider gun with two $B_1$'s}
\label{app:Nearmiss}

\begin{figure}[!htbp]
	\centering
	\includegraphics[width=0.8\linewidth]{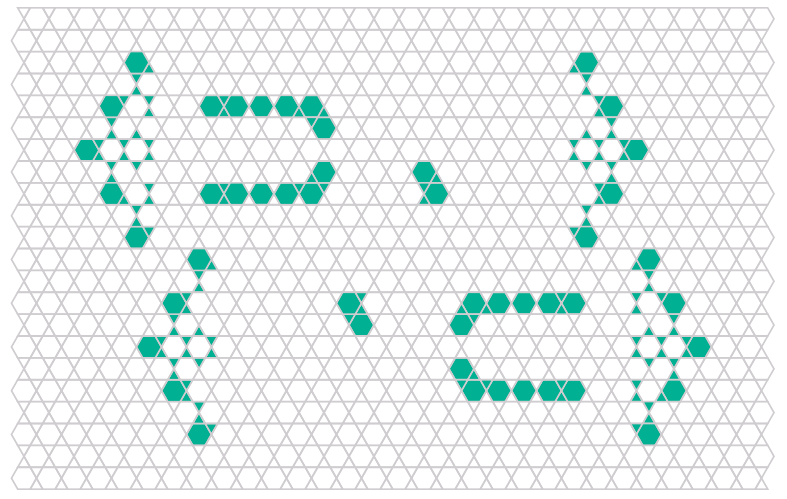}
    \caption{Near-miss glider gun configuration comprising two $B_1$ bouncers. Interactions between the intermediate $G$ glider trail produces two emitting $g$ gliders. The bounced $G$'s catch up with the emitted $g$'s and destroy the configuration just one step before they evade.}
	\label{fig:NearMissGliderGun}
\end{figure}

A snapshot of the near-miss glider gun solution comprising two $B_1$ bouncers described in Sec.~\ref{sec:GliderGun} is shown in Fig.~\ref{fig:NearMissGliderGun}.
\makeatletter
\def\bibsection{%
  \par
  \baselineskip26\p@
  \bib@device{\linewidth}{82\p@}%
  \nobreak\@nobreaktrue
  \addvspace{19\p@}%
  \par
}
\makeatother
\bibliography{Refs}
\end{document}